\documentclass[aps,prd,twocolumn,floatfix,nofootinbib]{revtex4}
\usepackage[pdftitle=Strings,pdfauthor=M.A. Fernandez]{hyperref}
\usepackage{graphicx,amsmath,amsfonts,amssymb,verbatim,float}
\begin{document}
\title{Cosmic Filaments from Cosmic Strings}
\author{M.A. Fernandez}\email{mfern027@ucr.edu}
\author{Simeon Bird}\email{sbird@ucr.edu}
\author{Yanou Cui}\email{yanou.cui@ucr.edu}
\affiliation{Department of Physics and Astronomy, University of California Riverside, 900 University Ave, Riverside, CA 92521}
\begin{abstract}
Cosmic strings are generically predicted in many extensions of the Standard Model of particle physics. We propose a new avenue for detecting cosmic strings through their effect on the filamentary structure in the cosmic web. Using cosmological simulations of the density wake from a cosmic string, we examine a variety of filament structure probes. We show that the largest effect of the cosmic string is an overdensity in the filament distribution around the string wake. The signal from the overdensity is stronger at higher redshift, and more robust with a wider field. We analyze the spatial distribution of filaments from a publicly available catalog of filaments built from SDSS galaxies. With existing data, we find no evidence for the presence of a cosmic string wake with string tension parameter $G\mu$ above $5\times 10^{-6}$. However, we project WFIRST will be able to detect a signal from such a wake at the $99\%$ confidence level at redshift $z=2$, with significantly higher confidence and the possibility of probing lower tensions ($G\mu \sim 10^{-6}$), at $z=10$. The sensitivity of this method is not competitive with constraints derived from the CMB. However, it provides an independent discovery channel at low redshift, which could be a smoking-gun in scenarios where the CMB bound can be weakened.
\end{abstract}

\maketitle

\section{Introduction} \label{introduction}

Many well-motivated fundamental physics theories beyond the Standard Model predict the existence of cosmic strings, which are approximately one-dimensional stable relic objects. They may arise in super-string theory as fundamental objects~\cite{Copeland:2003bj, Dvali:2003zj, Polchinski:2004ia, Jackson:2004zg, Tye:2005fn} or as vortex-like configurations of quantum fields such as those originated from a $U(1)$ symmetry breaking~\cite{Nielsen:1973cs, Kibble:1976sj}. A cosmic string network forms in the very early Universe, e.g., following a symmetry-breaking phase transition, and is expected to consist of stable horizon-length long strings together with sub-horizon loops that can decay away through gravitational radiation or matter emission \cite{Vilenkin:1981bx, Vachaspati:1984gt, Turok:1984cn, Burden:1985md, Olum:1999sg, Moore:2001px, Matsunami:2019fss, Vincent:1997cx, Bevis:2006mj, Cui:2008bd, Figueroa:2012kw, 1982PhLB..116..141K, 2018PhRvD..97j2002A}. The phenomenology of cosmic strings are characterized by their energy per unit length (tension) $\mu$ that relates to the symmetry breaking scale $\eta$ ($\mu\sim \eta^2$)~\cite{2000csot.book.....V}.

Cosmic strings have interested cosmologists and high energy theorists for decades, and can leave detectable signatures in a variety of observational windows. In the early days, strings were considered as potential large scale structure seeds \cite{1981PhRvD..24.2082V, 1981PhRvL..46.1169V}. This role was later ruled out by CMB data. Nevertheless, cosmic strings may still imprint the CMB as line discontinuities in the temperature map and the current constraint on the tension by Planck \cite{2014A&A...571A..25P} is $G\mu < 1.5\times 10^{-7}$ at $95\%$ confidence ($G$: Newton constant). A cosmic string background may produce detectable gravitational wave signatures that can reveal the expansion history of the early Universe as well as other beyond the Standard Model particle physics, and has thus received increasing attention in light of the recent LIGO detection~\cite{Cui:2017ufi, Cui:2018rwi, Abbott:2017mem, Auclair:2019wcv, Chang:2019mza, Gouttenoire:2019rtn, Gouttenoire:2019kij, Buchmuller:2019gfy, 2020arXiv200402889B, Dror:2019syi, Cui:2019kkd, 2018PhRvD..97j2002A}.

Cosmic strings can also be detected through lower redshift astrophysical observations, in particular through their distinct gravitational lensing effect which constrains $G\mu < 2.3\times 10^{-6}$ at $95\%$ confidence \cite{2010MNRAS.406.2452M}. Although these low $z$ lensing observations typically give weaker constraints than those based on the CMB or stochastic gravitational wave background (with model dependence), they serve as important complementary probes. Meanwhile, recent work has brought up the possibility that the CMB and stochastic gravitational wave background limits may be alleviated or evaded in certain scenarios, e.g., if the cosmic strings form before or during early stage of inflation and re-enter back into the horizon only at late times \cite{Guedes:2018afo, Cui:2019kkd}. In such cases low $z$ astrophysical signatures could be the smoking-gun for cosmic strings.

Early non-linearities are perhaps the most striking effect from cosmic strings in terms of structure formation. In particular, once inside the horizon, long strings straighten out at relativistic speeds \citep{2000csot.book.....V}. The spacetime metric around these straight segments is conic, with a deficit angle of the size $\sim 8\pi G\mu$ (i.e. one revolution around the string is less than $2\pi$ radians) \cite{1991PhST...36..114B}. This causes particles moving relative to the string to be kicked towards the plane traced out by the string, with a magnitude proportional to the deficit angle. A wake composed of in-falling particles is formed behind the string \cite{1984PhRvL..53.1700S, 1988CQGra...5...55D} and the wake grows according to standard linear theory \cite{1970A&A.....5...84Z}. Such an effect can alter the large scale structure by introducing non-linearities earlier in the universe than would otherwise be possible. 

Ultimately, these string induced non-linearities are overwhelmed by the growth of Gaussian fluctuations from inflation, erasing signatures of its existence at later times \cite{2016PhRvD..93l3501D}. Because of this erasure of the early structure signal, the CMB and lensing effects have been the more robust ways in which to constrain the tension of cosmic strings.

Here, we propose using the effect of the cosmic string on the filaments of the cosmic web as a new probe of cosmic strings. The large scale structure of the universe may be split into distinct components;  ``zero-dimensional'' halos at the intersections of ``one-dimensional'' filaments, which outline ``two-dimensional'' walls containing ``three-dimensional'' voids. These components, listed in descending order of density, make up the cosmic web \cite{1996Natur.380..603B}. The large scale cosmic web has been observed through the distribution of galaxies \cite{2001MNRAS.328.1039C, 2009MNRAS.399..683J, 2011AJ....142...72E, 2013ApJS..208....5N, 2015MNRAS.452.2087L, 2018A&A...609A..84S} and intensity maps of emission lines \cite{2010Natur.466..463C, 2018MNRAS.476.3382A, 2019BAAS...51c.101K}. Future proposed and planned intensity mapping surveys  \cite[e.g~][]{2015aska.confE..19S, 2020arXiv200205072C} and galaxy surveys such as those conducted by the square kilometre array \cite{2018arXiv181102743S}, or dark energy spectroscopic instrument \cite{2016arXiv161100036D} as well as the EUCLID \cite{2019arXiv191009273E} and WFIRST \cite{2015arXiv150303757S} satellites will extend our three-dimensional map of the cosmic web. Further boosting our interest in the large scale cosmic web is the recent development of a multitude of cosmic web identification algorithms \cite{2018MNRAS.473.1195L}, some of which have already identified filaments in existing galaxy surveys \cite{2014MNRAS.438.3465T, 2020arXiv200201486M}.

In this paper we look at the effect that the passage of a long, straight cosmic string has on the filamentary structure in dark matter (DM) simulations. We examine filaments here for two reasons. First, while the wake is effectively a two dimensional feature, when projected onto the sky it would appear one dimensional. Second, filament finders are well developed and may be robustly applied to galaxy catalogues. It may be possible to detect the effects of a cosmic string in other cosmic web structures, such as walls and voids, but we defer that to future work. We consider string tensions ranging from those consistent with Planck limits to a value similar to the gravitational lensing limit. This method is sensitive to higher redshift strings (when the string has passed through the simulation volume and the wake forms, $z \sim 100$) and larger scales ($\sim10$ Mpc) than the lensing signal ($z \lesssim 1$, $\sim$Mpc). 

Refs.~\cite{2018arXiv180709820L, 2018arXiv181007737C, 2018arXiv180400083C} showed that tensions of $G\mu\sim10^{-7}$ are detectable in cosmological simulations using the cosmic string wake. However, these methods require an accurate map of the three-dimensional dark matter distribution at $z > 2$, which is currently out of reach of foreseeable experiments. Another potential, albeit at present futuristic, method for detecting the wake from cosmic strings is 21 cm intensity mapping \cite{2010JCAP...12..028B}. The sensitivities using the method in our paper are weaker (down to $G\mu\sim10^{-6}$), but can be achieved with near future galaxy surveys such as WFIRST and EUCLID.

To the authors' knowledge, the effect of the physical shift and kick due to the string on the statistics of cosmic filaments has not been explored. While we will show that the effects are generally too small to be competitive with limits from CMB constraints, the novel independent method we propose provides a valuable complementary probe, and would become competitive in scenarios when CMB bounds are weakened (for example, regrowing cosmic strings after inflationary dilution \cite{Cui:2019kkd}).

In Section~\ref{simulations} we detail the suite of simulations that we will be using throughout, as well as the method we employ to add the effects of the cosmic string to the simulation. We also comment on the possibility of detection via the kinematics of the halos, as well the velocity kick feature of the wake implementation. In Section~\ref{filaments} we describe the two filament identification algorithms we use, comparing the filaments they identify qualitatively. The result of this section is a catalog of filaments that we analyze in Section \ref{results}, along with a set of publicly available filaments constructed from SDSS galaxies. We also demonstrate the projected sensitivities for WFIRST in Section \ref{results}. Finally, in Section \ref{conclusion} we conclude, pointing to future prospects for cosmic string wake detection. In Appendix \ref{boxsize} we show that our results are independent of the box size and resolution of our simulations.

\section{Simulations} \label{simulations}

We performed a suite of dark matter only cosmological simulations using MP-Gadget\footnotemark[1] \cite{2019JCAP...02..050B}. MP-Gadget is a fork of Gadget-3 \cite{2005MNRAS.364.1105S} modified for scalability. The initial power spectrum is generated via the Boltzmann code CLASS \cite{2011arXiv1104.2932L}. Radiation density is included in the background expansion rate, and the simulation box has periodic boundaries. The cosmology parameters are the current defaults of MP-Gadget and are consistent with the nine-year WMAP results \cite{2013ApJS..208...19H} (our results are not sensitive to the exact values used). These, and the simulation settings used, can be seen in Table \ref{table:params}.
\footnotetext[1]{\url{https://github.com/sbird/MP-Gadget3}}

\begin{table}
\caption{\label{table:params} Parameters}
\begin{ruledtabular}
\begin{tabular}{cc|cc}
Cosmology\footnotemark[1] & & Simulation\footnotemark[2]\\
\hline
$\Omega_0$         & $0.2814$ & $N$               & $512^3$\\
$\Omega_{\Lambda}$ & $0.7186$ & $L$ \text{[Mpc/h]}   & $64$ \\
$\Omega_b$         & $0.0464$ &      $\ell$ \text{[Mpc/h]}        & $250$\\
$H_0$ \text{[km/s/Mpc]}       & $69.7$   &     $z_0$       & $99$\\
$\sigma_8$         & $0.810$  &     $z_{cs}$       & $31$\\
$n_s$              & $0.971$  &            & 
\end{tabular}
\end{ruledtabular}
\footnotetext[1]{Total matter $\Omega_0$, dark energy $\Omega_{\Lambda}$, and baryon $\Omega_b$ densities. Hubble constant $H_0$, density fluctuation $\sigma_8$, scalar spectral index $n_s$.}
\footnotetext[2]{Number of DM particles $N$, box length for main simulations $L$, box length for scaling simulation $\ell$ (see Appendix \ref{boxsize}), initial redshift $z_0$, wake insertion redshift $z_{cs}$.}
\end{table}

The main results of this paper are from a set of five simulations (each with a different starting random seed, leading to distinct dark matter structures) with the parameters in Table \ref{table:params} and a mass resolution of $1.5\times 10^{8} M_{\odot}$. The effects of the string wake were inserted at $z=31$. The insertion redshift is chosen to be late enough that a string will have time to travel the box length since matter-radiation equality and early enough that non-linear structure has not yet formed in the wake. The insertion time does not affect our results as long as it is early enough that linear perturbation theory (Equations \ref{eqs:shift} \& \ref{eqs:kick}) is a good description. This will be until structure formation dominates the wake signal \cite{2016PhRvD..93l3501D}. An additional simulation with a larger box size ($250$ Mpc) was run to ensure that our results were not strongly affected by box size or resolution (see Appendix \ref{boxsize}).

\subsection{Cosmic String Wakes} \label{wakes}
\begin{figure}
\centering
\includegraphics[width=0.4\textwidth]{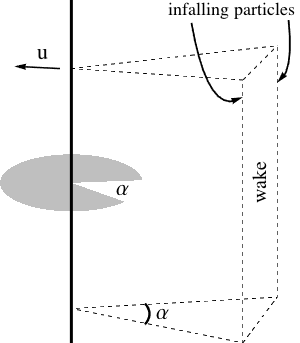}
\caption{\emph{Setup for cosmic string \& wake.} The cosmic string (bold, black) creates a deficit angle, $8\pi G\mu$, in the space around it. The string travels with speed $u$, leaving behind a wedge shaped overdensity called a wake with opening angle, $8\pi G\mu\gamma$, where $\gamma=(1-u^2)^{-1/2}$.}
\label{fig:cs_cartoon}
\end{figure}

Cosmic strings create a deficit angle in the spacetime surrounding them. This means that as a cosmic string passes through matter, the trajectory of that matter is altered (for more details, see \cite{1991PhST...36..114B} or \cite{2000csot.book.....V}). Specifically, the matter is ``kicked'' towards the worldsheet of the string with an additional velocity given by
\begin{equation}
    \delta v = 4\pi G\mu \gamma u,
\end{equation}

where $G\mu$ is the dimensionless parameter defined by the tension of the cosmic string, $u$ is the comoving speed of the cosmic string, and $\gamma$ is the usual relativistic Lorentz factor introduced by transforming to the reference frame of the particle \cite{2000csot.book.....V}. See Figure \ref{fig:cs_cartoon} for a cartoon of the cosmic string wake, deficit angle, and wake. We set the string speed as $u = 0.6c$, consistent with e.g., \cite{Cui:2018rwi} (Eqs. 2.3 \& 2.4 therein). The string speed only affects the magnitude of the velocity kick and displacement of particles on wake insertion. Therefore, the exact value is sub-dominant compared to the tension parameter, which in this study varies by a factor of $50$, as compared with possible string speeds varying from the value used here on the order of a factor of $\sim 1-5$ \cite{2000csot.book.....V}.

These in-falling particles create an overdensity along the worldsheet of the cosmic string. This overdensity, or cosmic string ``wake'', is an early non-linear structure which perturbs the position and velocity of nearby particles. Instead of simulating the cosmic string directly, the wake is simulated following the method used in \cite{2018arXiv180400083C}. We simulate the wake once the string has passed by including the linear density perturbation induced by the string's passage. For simulation purposes, a snapshot that is sufficiently late that the string will have passed through the box is chosen, and the particles are displaced and kicked at that time (by an amount corresponding to the prediction of linear theory between the time of the string's passage and the time of wake insertion).

Perturbations due to the string wakes are inserted as
\begin{align}
    \delta x &= \frac{3}{5}\delta v \ t_{eq}\frac{(z_{eq}+1)^2}{(z+1)}h,\label{eqs:shift}\\
    \delta \dot{x} &= \frac{2}{5} \delta v \ \frac{t_{eq}}{t}\frac{(z_{eq}+1)^2}{(z+1)^2}, \label{eqs:kick}
\end{align}

where $t_{eq}, z_{eq}$ are the time and redshift of matter-radiation equality (and when the wake is first formed), $h$ is the Hubble factor, and $t, z$ are the time and redshift at which the shift and kick are calculated (i.e. when the wake effect is inserted into the simulation). Note that our perturbations differ slightly from \cite{2018arXiv180400083C}: the factor of $h$ in the shift is due to positions in MP-Gadget being in comoving kpc/h, one factor of $z^{-1}$ in the velocity kick is due to MP-Gadget using physical peculiar velocities, and we have dropped the (good) approximation that $z \approx z+1 = 1/a$.

In practice, the snapshot at which the wake's effects are being inserted is adjusted by first shifting the particle positions along the coordinate perpendicular to the plane of the string wake. The particles on either side are shifted towards the plane, then the velocity perturbation is added towards the plane of the wake. The simulation is then run to completion from the updated snapshot, which now includes the effects of the wake. Given that we expect $\sim10$ horizon-length strings per horizon \cite{2015SchpJ..1031682V, 2011PhRvD..83h3514B}, we would not expect to see multiple string wakes in a region the size of our simulation boxes, therefore only one wake is inserted in each simulation.

\begin{figure}
\centering
\includegraphics[width=0.45\textwidth]{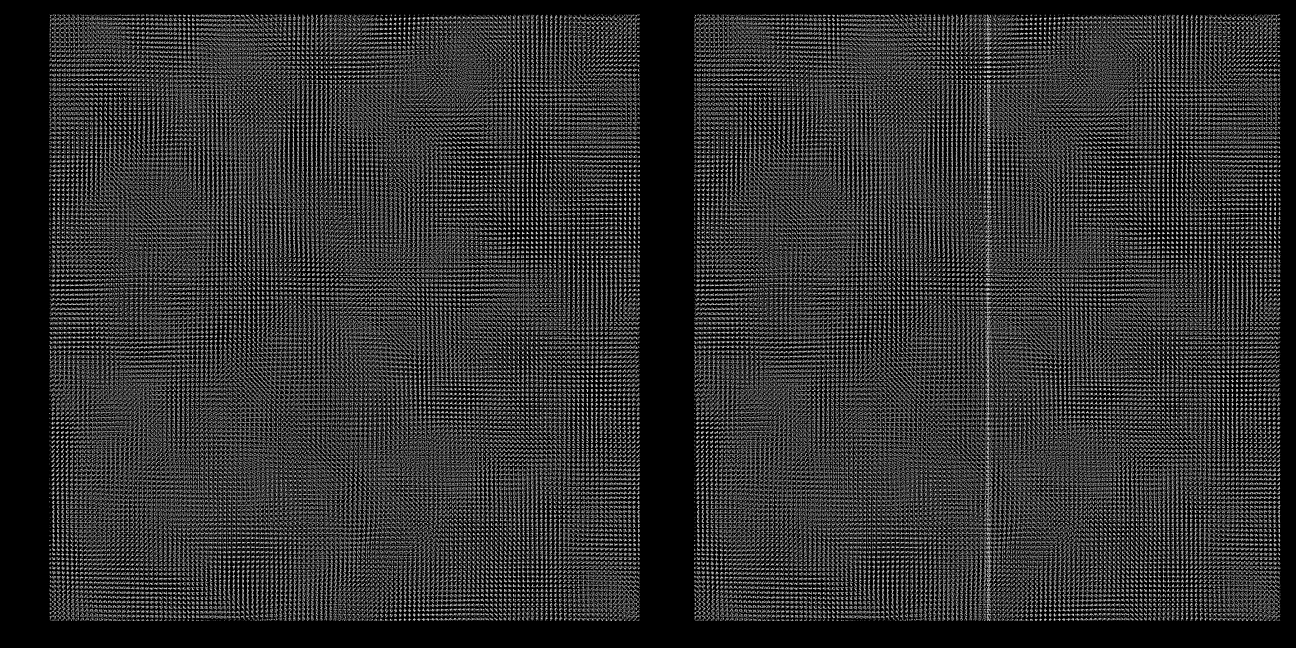}
\includegraphics[width=0.45\textwidth]{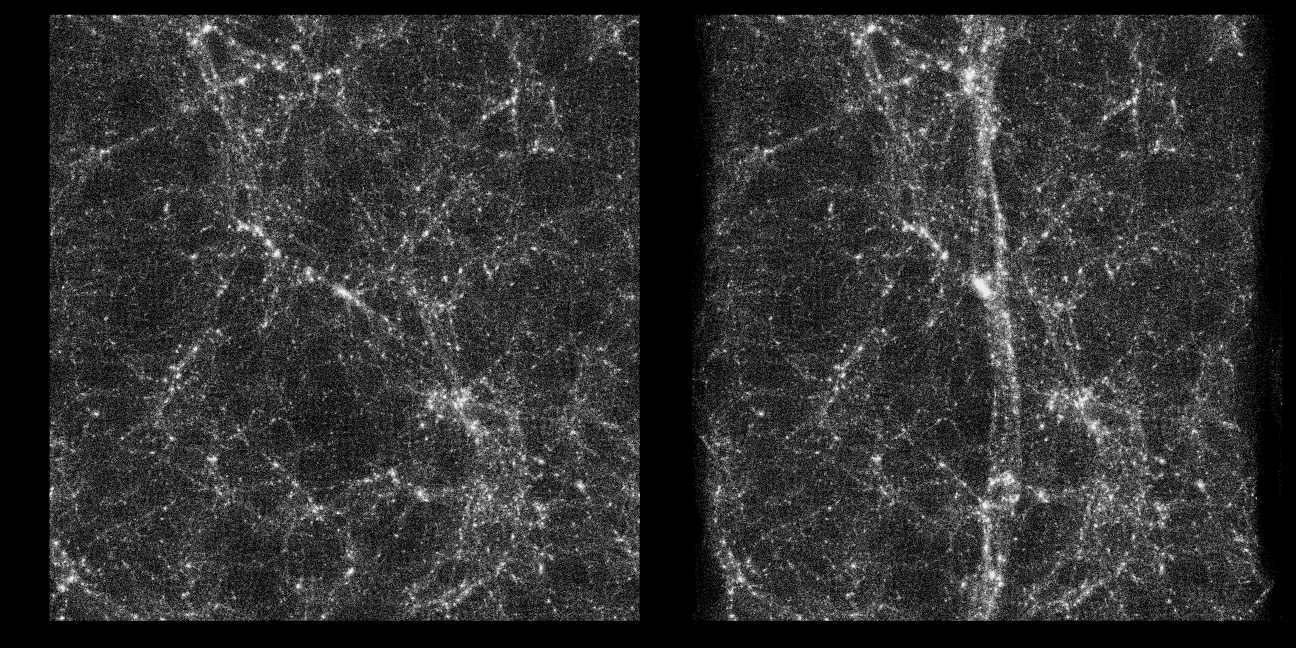}
\caption{\emph{Visual wake versus no-wake comparison.} Projection of the particle density at $z=31$ (top) and $z=1$ (bottom). Left-hand panels are from a simulation with no wake, right-hand panels are from a simulation with a $G\mu = 4 \times 10^{-6}$ wake inserted at $z=31$.}
\label{fig:sim_snaps}
\end{figure}

The wake insertion is performed in the same way for each of the five distinct simulated cosmic structures. For each of these distinct structures, four different scenarios are simulated: one scenario where there is no wake, and the remaining scenarios where a wake is inserted with tensions $G\mu = 10^{-7}, 10^{-6}$, and $5\times10^{-6}$. An example can be seen in Figure \ref{fig:sim_snaps}, which shows the density of DM particles at two redshifts for the case with no wake (left) and with a $G\mu = 4\times10^{-6}$ wake (right). 

\subsection{Velocity} \label{velocity}

The effect of the cosmic string passage is a velocity kick towards the worldsheet of the string. This leads to an effect on the velocity distribution of the halos. At early times there is an excess velocity towards the wake. At some later point, depending on the string tension, the kicked halos cross the wake and the excess velocity is away from it. While this is a significant noticeable effect on the redshift space distribution of the halos, it depends strongly on the orientation between the observer and wake. Our simulations indicate that while even the lowest tension string wake would have a noticeable signal down to $z=2$, the signal disappears when the angle between observer and the plane perpendicular to the wake is greater than $\sim10$ degrees at higher redshift ($z \ge 5$) and $\sim3$ degrees at lower ($z \le 2$). A detection via this signal would rely on a highly fortuitous alignment.

\begin{figure}
\centering
\includegraphics[width=0.45\textwidth]{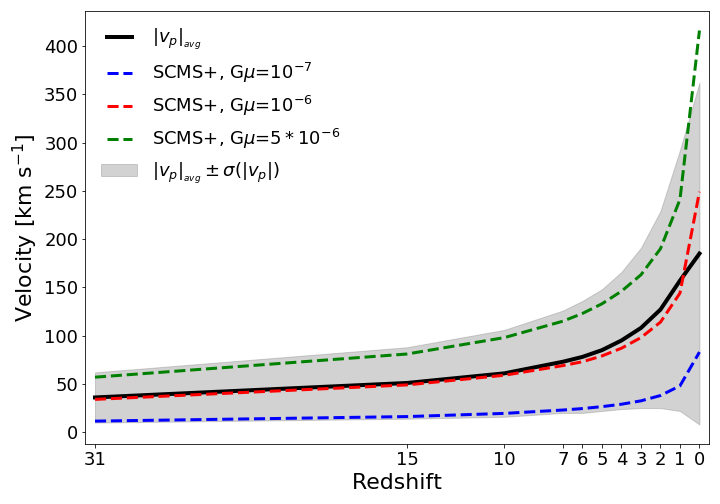}
\caption{\emph{Average particle speed perpendicular to wake.} Average speed of DM particles in the direction perpendicular to the cosmic string wake. Dashed lines show the kick imparted on the particles (see Section \ref{wakes}) for the three different tensions used throughout.}
\label{fig:vel_kick_v_avg_vel}
\end{figure}

These signals depend on the ratio of the average speed of particles towards the wake prior to wake insertion, and the velocity kick they receive. Figure \ref{fig:vel_kick_v_avg_vel} shows the average speed perpendicular to the wake against redshift. Also shown is the $1\sigma$ region of these speeds (shaded). The velocity kicks from wake insertion are shown as dashed lines. The middle tension case traces out the particle speed fairly well, while the other two tensions trace out the edges of the $1\sigma$ region. As might be expected, the observability of the effect from the passage of a cosmic string seems to depend on how large a kick the particles receive compared to the intrinsic velocity dispersion from standard DM structure formation (i.e. a non-trivial number of halos change direction due to the wake).

\section{Filament Identification} \label{filaments}

While the cosmic string wake is inserted via a position and velocity perturbation on the DM \textit{particles}, we identify the filamentary structure from the resulting DM \textit{halos} (these are constructed using a friends-of-friends algorithm in the simulation code). This is much more computationally reasonable, and agrees more closely with how filamentary structure is currently identified observationally (i.e. via galaxy surveys).

Many algorithms have been devised to identify cosmic web components within cosmological simulations or from galaxy catalogues (see \cite{2018MNRAS.473.1195L} for a comparison of 12 algorithms). Here, we use two such algorithms; the subspace constrained mean shift (SCMS) algorithm \cite{2011jmlr.12.1249, 2015MNRAS.454.1140C}, and DisPerSE (Discrete Persistent Structures Extractor) \cite{2011MNRAS.414..350S, 2011MNRAS.414..384S}. In both cases we identify only linear structures, the filaments of the cosmic web. By using two methods we mitigate the dependence of our results on the identification algorithm used, or options/parameters selected within each method. The two selected methods are at somewhat opposite ends in terms of complexity (with SCMS being fairly intuitive), and implementation (DisPerSE is a downloadable installation).

Below we briefly describe each algorithm and note our parameter/option choices in using them, as well as any necessary alterations made to conform to our simulation output (the most notable being the periodicity of our simulation box).

\subsection{SCMS+} \label{scms}

\begin{figure}
\centering
\includegraphics[width=0.45\textwidth]{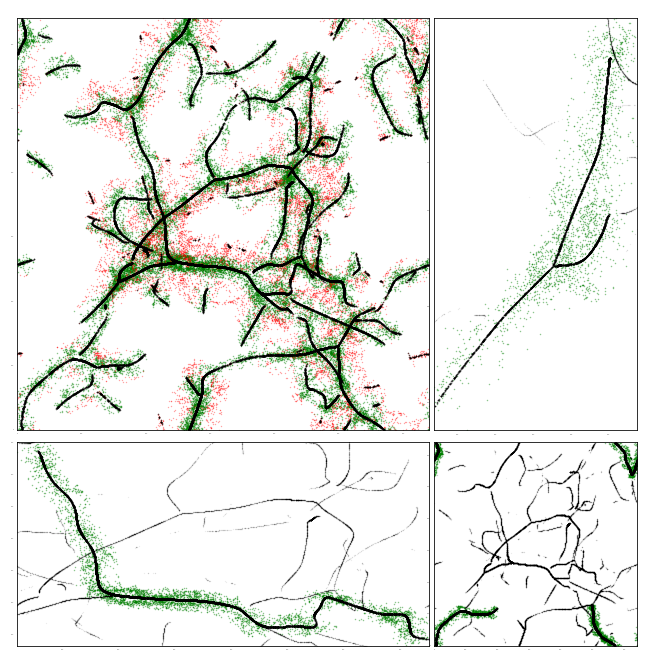}
\caption{\emph{SCMS+ filaments.} (top left) converged SCMS tracers (black) outlining the spine of the filaments, halos-in-filaments (green), and halos too far from the spine (red). (top right) a split filament. (bottom right) a filament straddling the periodic boundary of the box. (bottom left) a ``wiggly'' filament.}
\label{fig:segment_plot}
\end{figure}

SCMS is a gradient ascent method, which shifts tracer particles according to the Hessian (second derivative, or curvature) matrix of the density at each tracer position. The density is approximated with a Gaussian kernel density estimate (KDE) at each tracer position according to the population of dark matter halos in the simulation box. We follow the implementation as described in \cite{2015MNRAS.454.1140C}, which thresholds out halos in low density regions, then runs the SCMS algorithm to identify ridges in the KDE. Here, we use the DM halos as the starting positions of the tracers and set the smoothing length as 2 Mpc. Some modifications were implemented to correctly handle the periodicity of the simulation box. A brief outline of the algorithm follows.

\begin{itemize}
    \item (Thresholding) Tracers in low density regions are removed by calculating the KDE for all tracers and removing those with values lower than the mean KDE, as these are unlikely to reside in filaments (\cite{2018MNRAS.473.1195L} figures 4 \& 5). Particle separations account for the box periodicity, connecting particles using the smallest possible great circle arc.
    \item (SCMS) Using the subset left over from the thresholding step, the SCMS algorithm runs until a tracer is shifted by less than 1 kpc. For each iteration the Hessian matrix is calculated and the smallest two eigenvectors dictate which direction the halo moves to ascend the local ridge. Once a tracer is converged the next tracer is shifted until convergence, and so on (the KDE calculated here is from the DM halos, which are not shifted). To accommodate the periodic boundaries, the box (and the halos within) is centered on the current tracer position, and the separations are calculated between this tracer and tracer-centered halos. At the end of each step, if the tracer has shifted outside the box, it is moved back to the correct, periodic position by adding or subtracting the box size.
\end{itemize}

The output of the SCMS algorithm is a set of tracers converged on the ridges of the density field, and is not a set of filaments. To obtain separated filaments (and filament properties) we use both the tracers and the DM halos. First, the tracers are connected into segments if they are separated by less than a maximum distance, with a value that depends on the redshift. The segments are then connected iteratively using the same, redshift dependent maximum distance (from $250$ kpc/h at $z=0$ to $1200$ kpc/h at $z=10$) until a set of separated filaments is achieved. We call the entire method, from thresholding to filament separation, SCMS+. Some examples of filaments identified with this algorithm can be seen in Figure \ref{fig:segment_plot}, which for a simulation with no wake inserted, shows the entire box in the top left corner (converged tracers in black, halos belonging to filaments in green, halos too far from filaments in red), then some zoomed in examples of filaments in the other panels.

The length of the filaments are determined by a smoothed stepping procedure, which sums the distances between neighboring tracers, which have been smoothed by 0.5 Mpc, starting from one end of each filament (a filament end is determined by identifying the tracer with the fewest directions occupied by another tracer). DM halos are then assigned to filaments based on proximity and a cutoff of 2 Mpc. Filaments that are shorter than the smoothing length (2 Mpc) or have fewer than 3 halos assigned to them are discarded. Finally, the mass of each filament is the sum of the halos assigned to that filament.

\subsection{DisPerSE} \label{disperse}

\begin{figure}
\centering
\includegraphics[width=0.4\textwidth]{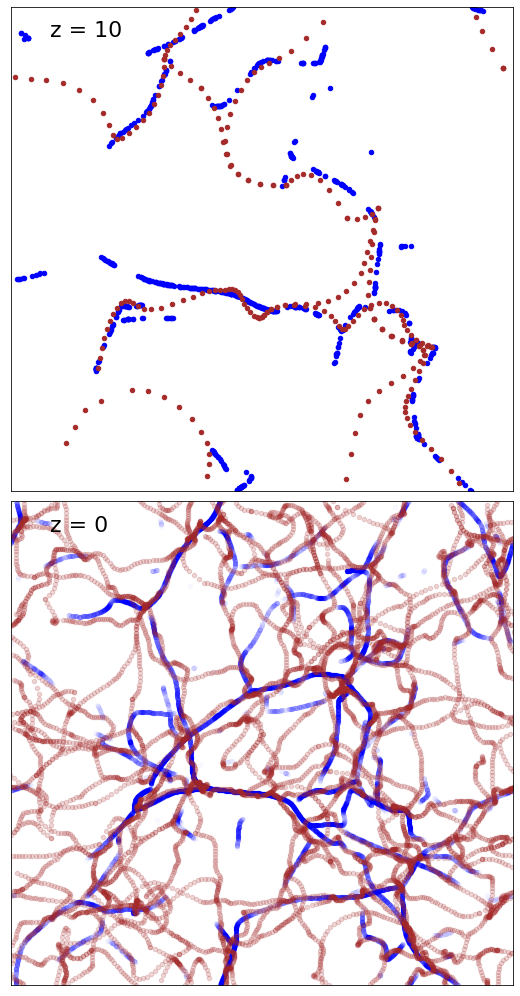}
\caption{\emph{Visually comparing SCMS+ to DisPerSE.} SCMS+ filaments (blue) and DisPerSE (brown) filaments, both using DM halos as the input. The top panel is at $z=10$, while the bottom is at $z=0$. While DisPerSE is clearly more sensitive, finding more filaments, the largest filaments are similar.}
\label{fig:disperse_v_scms}
\end{figure}

DisPerSE is a widely used cosmic web identification tool which extracts structure using the Morse-Smale complex of the input. The input is the Delaunay tessellation of the particle distribution. The details of DisPerSE are outside the scope of this work (an overview and tutorial are available\footnotemark[2]\footnotetext[2]{\url{http://www2.iap.fr/users/sousbie/web/html/indexd41d.html}}). We implement the main program of DisPerSE, MSE, on the tessellated DM halo population from our simulations with a 6$\sigma$ persistence threshold. The output is converted using the skelconv program in DisPerSE, smoothed over 10 halos, and filaments are assembled if the angle between them is less than 75$^{\circ}$.

DisPerSE can find filaments directly from DM particles\footnotemark[3]\footnotetext[3]{In principle, so can SCMS, but it scales poorly to large tracer numbers.}. To check the robustness of our results we subsample 5\% of the DM particles from the simulations and run DisPerSE using the same settings, except for increasing the smoothing to 100 particles. We visually inspected the output and confirmed that approximately three-quarters of the filaments found from halos had approximately the same positions as the filaments found from particles. For ease of comparison with SCMS+ filaments, we use DM halos for the remainder of this work.

\begin{figure}
\centering
\includegraphics[width=0.45\textwidth]{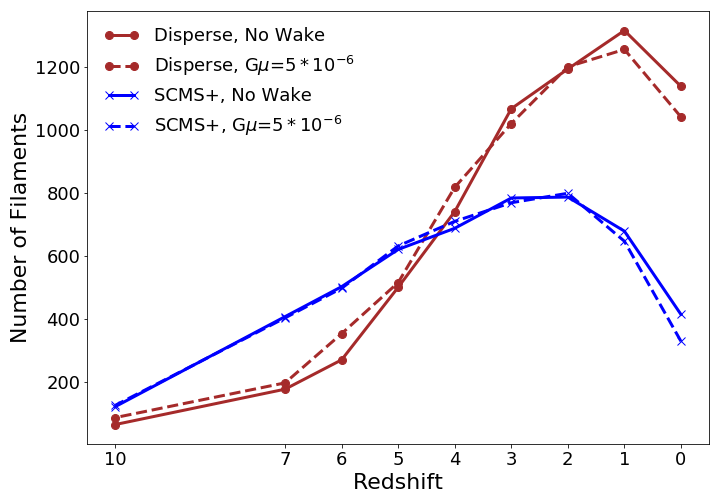}
\caption{\emph{Filament number comparison (SCMS+, DisPerSE).} Number of filaments identified by the two methods as a function of redshift. Included are the filaments from the simulation without a wake, and a simulation with a $G\mu = 5\times10^{-6}$ wake.}
\label{fig:nfil_v_redshift_number}
\end{figure}

While the two filament identification algorithms employ fairly different methods, they both identify similar major filaments. DisPerSE detects substantially more filamentary structure as can be seen in Figure \ref{fig:disperse_v_scms}, which shows the filaments identified from a simulation with no wake. Figure \ref{fig:nfil_v_redshift_number}~compares the number of filaments identified by each method as the simulation evolves. Both algorithms detect the growth of structure up to the onset of dark energy domination and a subsequent decline in filament number. However, the two methods differ in filament number and redshift trend: DisPerSE initially has fewer filaments but much larger growth with redshift, which peaks at a higher level.

Both the number of filaments and the trend with redshift depend strongly on the values chosen for key parameters for each method; for DisPerSE the persistence threshold, and for SCMS+ the smoothing length. For example, reducing the persistence threshold of DisPerSE from $6\sigma$ to $2\sigma$ leads to a $\sim 10$-fold increase in filaments at all redshifts and moves the peak redshift to $z=2$, preserving the late decline in filament number. Despite the differences in the filament populations these methods return, neither method shows a convincing signature of a cosmic string wake on structure formation. We also stress that in the analysis presented in this paper, we are concerned with relative changes, comparing similar simulations where the only difference is the inclusion of a cosmic string wake.

For each cosmic string tension we have performed five different simulations using different initial realisations of cosmic structure. For each string tension, the filament catalogs from these five different structure realisations are combined, boosting the sample of available filaments. The final results are two catalogs (SCMS+, DisPerSE) with four filament populations each\footnotemark[4]\footnotetext[4]{(python) code used to produce these catalogs are available at \url{https://github.com/mafern/SCMSplus}.}. In Section~\ref{results} we analyze these populations for potential cosmic string signals, focusing on the SCMS+ catalog.

\section{Results} \label{results}

Our goal at the outset of this paper is to determine if there are signals in the filament population indicating the previous passage of a cosmic string. With a catalog of filaments in hand, we compare the filament population and spatial distribution of filaments in simulations with and without cosmic string wakes.

\subsection{Filament Population} \label{number_properties}

\begin{figure}
\centering
\includegraphics[width=0.45\textwidth]{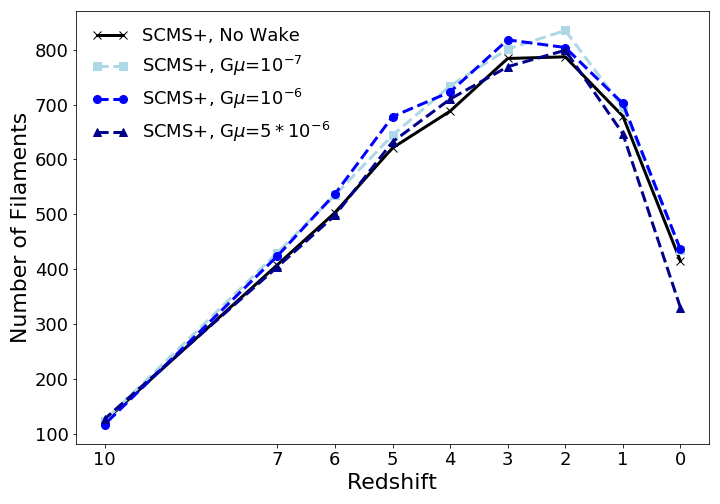}
\caption{\emph{Filament number, increasing tension (SCMS+).} Number of filaments identified by SCMS+ versus redshift. Even at very large tension, the effect on the number of filaments is small and noise dominated.}
\label{fig:nfil_tension_scms}
\end{figure}

Figure \ref{fig:nfil_tension_scms} shows the number of filaments for each of the four wake scenarios, plotted against redshift for the results from the SCMS+ filaments. Similar results are obtained with DisPerSE.  The change in the number of filaments between the simulation without a wake and the three with wakes is small and inconsistent. It appears that the presence of the wake both separates and connects filament segments, in approximately equal amounts. This leads to little to no change in the number of filaments.

\begin{figure*}
\centering
\includegraphics[width=0.80\textwidth]{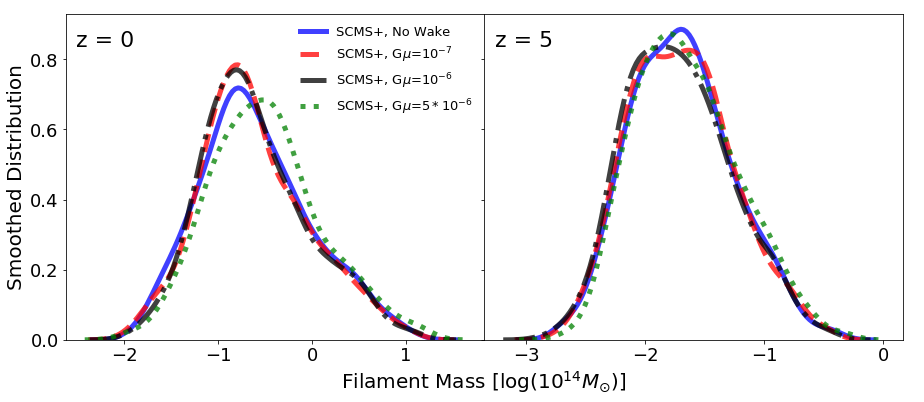}
\caption{\emph{SCMS+ filament mass distributions.} Smoothed, normalized distribution of filament masses at $z=0$ (left) and $z=5$ (right), for the three wake and no-wake cases. There are no significant changes in the masses of the filament population early or late in the simulation.}
\label{fig:mass_dist}
\end{figure*}

The properties of the filament populations, for example the distribution of filament masses, are another potential signal. The procedure for determining the filament length and mass are outlined at the end of Section \ref{scms}. As can be seen in Figure \ref{fig:mass_dist}, which shows the distribution of filament masses for the SCMS+ filaments, the cosmic string wake has very little effect on the filament masses. The distribution of filament lengths is similarly devoid of a distinguishing signal. Visual inspection of the filament distribution reveals, however, that the cosmic string wake does alter the spatial distribution of filaments, enlarging some and disrupting others. Once averaged over the whole box, the overall distribution is unchanged. Filaments are re-ordered in a way which is indistinguishable from the random variance due to the realization of structure formation. One feature of structure resulting from the wake which distinguishes it from structure formation seeded from inflationary perturbations is that it has a preferred direction (i.e. towards the wake). This is the feature that we exploit in the following section.

\subsection{Spatial Distribution of Filaments} \label{spatial_distribution}

\begin{figure}
\centering
\includegraphics[width=0.45\textwidth]{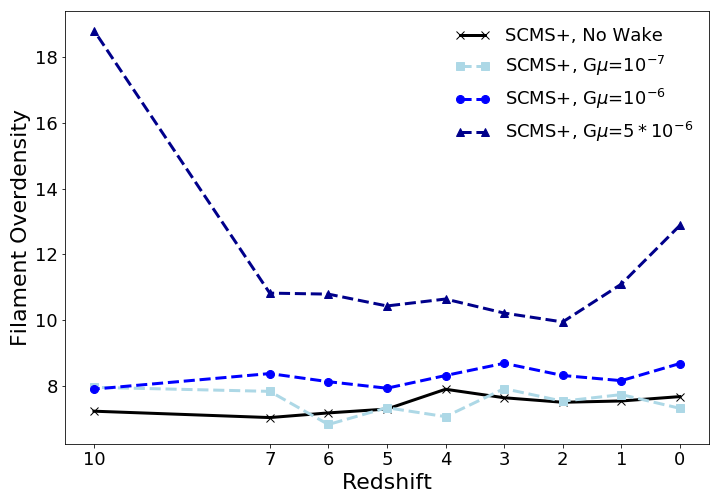}
\caption{\emph{Central filament overdensity, increasing tension.} Overdensity of filaments identified by SCMS+ versus redshift. The overdensity becomes more pronounced as the tension of the string is increased.}
\label{fig:od_tension_scms}
\end{figure}

Now we illustrate more sensitive observables for identifying filaments originated from string wakes. The effect of a cosmic string is to pull matter towards the worldsheet of its passage, motivating a look at the spatial distribution of filaments. In principle the spatial distribution of two-dimensional walls and three-dimensional voids will be affected similarly, however we remain focused on filaments here. We do this in two ways, the first being a comparison of the filament number density around the center of the wake to the overall filament number density. This is shown in Figure \ref{fig:od_tension_scms} for the SCMS+ catalog (we obtained similar results for the DisPerSE filaments). Any part of a filament within the central $10$ Mpc/h is added to the number of filaments near the wake. The number density in this central region, surrounding the wake, is divided by the filament number density for the whole box. As a baseline the same calculation is repeated for simulations with no wake, comparing the central region to the rest of the box.

The smallest tension wake we include here ($G\mu = 10^{-7}$) is not distinguishable from the simulation without a wake. The intermediate tension wake ($G\mu = 10^{-6}$) shows signs at higher redshift ($z > 5$) of a central overdensity in the number of filaments. The largest tension wake we include ($G\mu = 5 \times 10^{-6}$) shows a clear overdensity in the central region at all redshifts, with the signal generally decreasing with time, with the exception of an uptick after redshift $2$. This is likely due to the filaments coalescing into larger structures, which already have segments within the central region, see Figure \ref{fig:nfil_tension_scms}). We do not see this trend in the lower tension and no-wake simulations because in these cases the filaments have not been significantly distorted/pulled towards the central wake.

In an observational setting, we would not know a priori where the wake is, and therefore the previous comparison would be of limited use. Rather than looking only at a slice of the box parallel to and around the wake center, we can look at slices centered from one side of the box to the other. The number of filaments in each $10$ Mpc/h slice can then be compared to the mean number of filaments in a slice, i.e. each slice filament number is divided by the mean slice filament number for the entire box. A value higher than one is overdense with respect to the mean slice. Comparing the overdensity in these slices highlights any clustering of filaments around a potential wake.

\begin{figure*}
\centering
\includegraphics[width=0.80\textwidth]{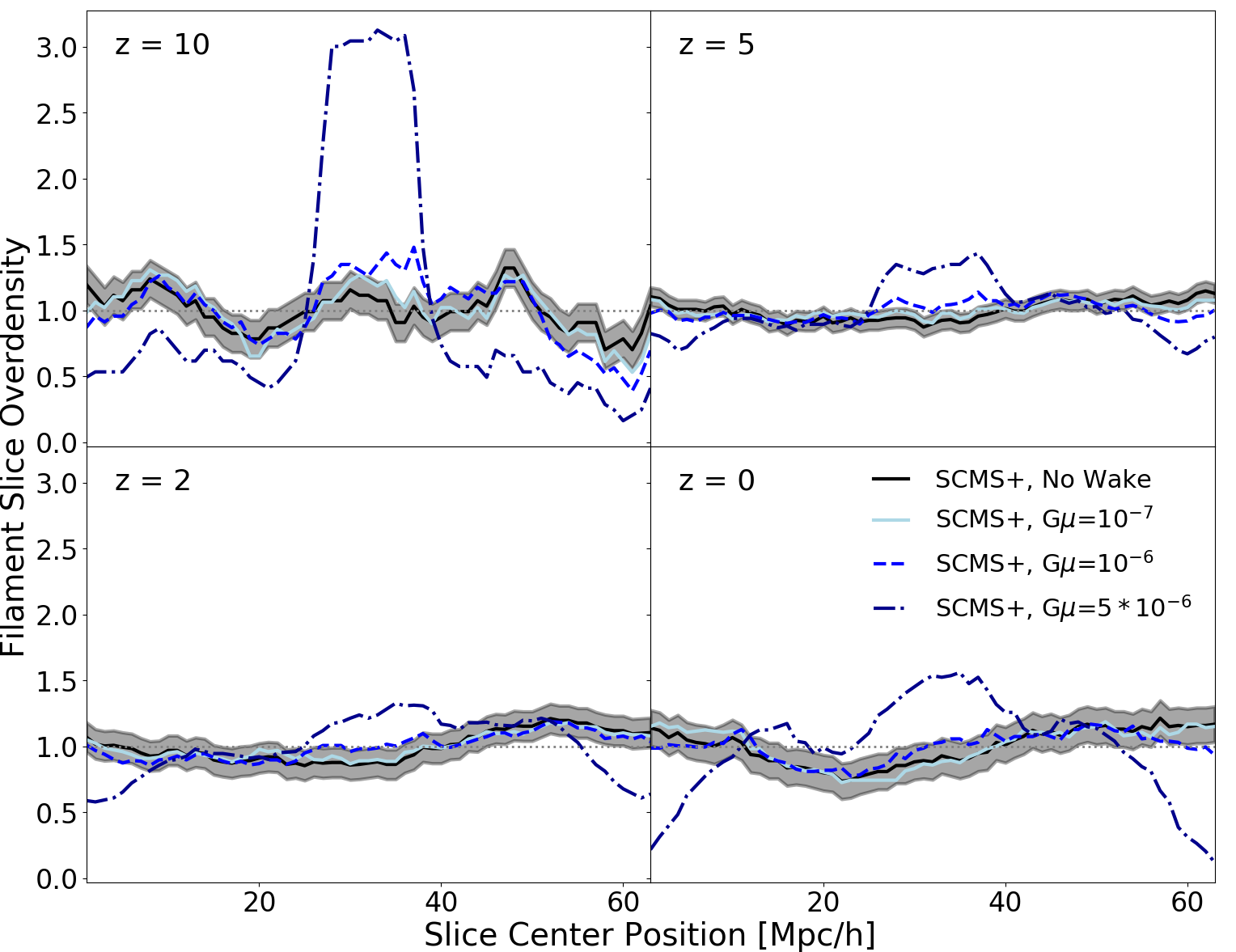}
\caption{\emph{Filament slice overdensity (SCMS+).} Overdensity of filaments identified by the SCMS+ method in $10$ Mpc/h slices versus the center of the slice. The grey shaded region shows the $1\sigma$ area for the no wake case.}
\label{fig:slice_tension_scms_w5}
\end{figure*}

In Figure \ref{fig:slice_tension_scms_w5} the \textit{slice overdensity} is apparent in the center of the box at all redshifts for the largest tension wake, while the smaller tension wakes show no robust signal at any redshift. The signal from the DisPerSE filaments is marginally smaller, but still shows that only the largest tension wake produces a noticeable overdensity. The underdensity at the edges of each panel is due to the finite size of the box. During wake insertion particles are kicked towards the center of the wake, away from the edges \cite{2018arXiv180400083C}.

\begin{figure}
\centering
\includegraphics[width=0.45\textwidth]{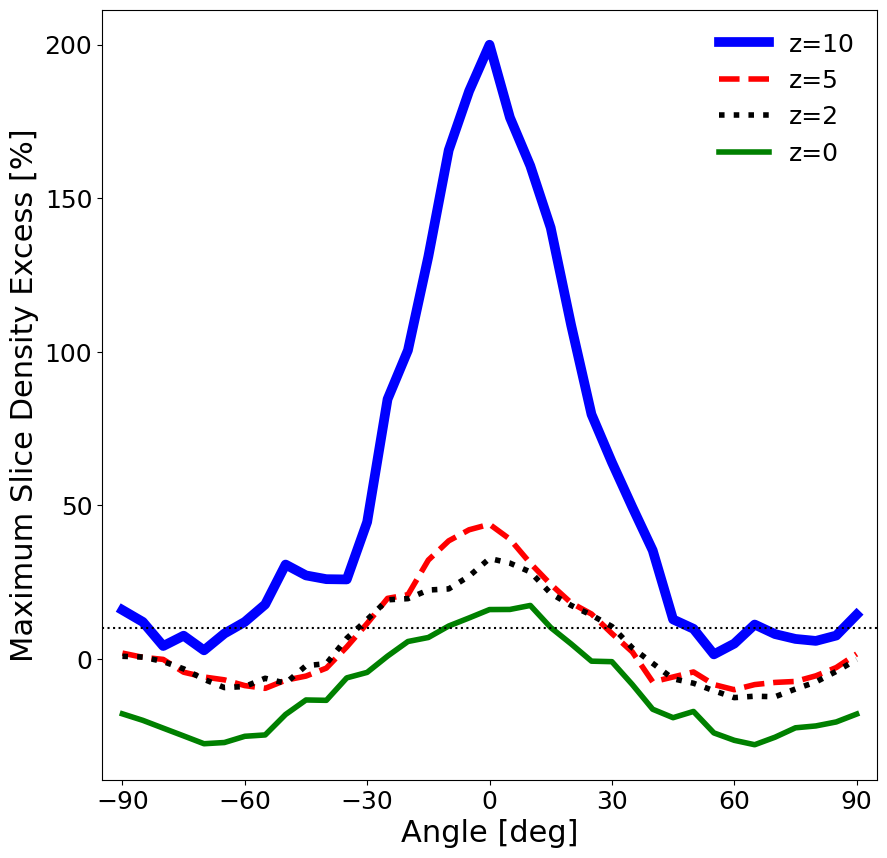}
\caption{\emph{Maximum number density excess by viewing angle.} The maximum percent difference between the number of filaments in a slice for the simulation with a $G\mu=5\times10^{-6}$ tension wake to the $1\sigma$ region in the no-wake simulation (see grey shaded region in Figure \ref{fig:slice_tension_scms_w5}). The dotted line marks a $10\%$ excess.}
\label{fig:slice_angle}
\end{figure}

\begin{figure}
\centering
\includegraphics[width=0.45\textwidth]{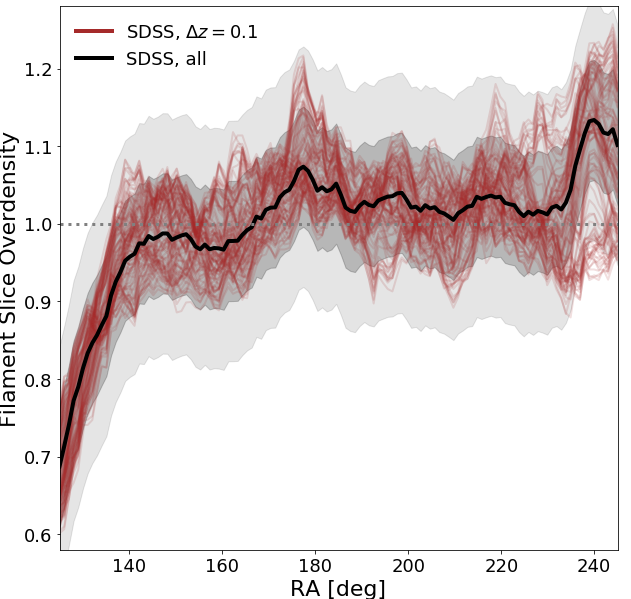}
\caption{\emph{Filament slice overdensity (SDSS).} Overdensity of SDSS filaments constructed using the filament catalog from \cite{2016MNRAS.461.3896C}. Slices span $10$ degrees in RA and are plotted over the RA slice center. Shown are slice overdensities using filaments at all redshifts (black), the $1\sigma$ and $2\sigma$ regions (dark and light shaded), and each $\Delta z = 0.1$ bin from $z=0.005$ to $z=0.7$ (brown).}
\label{fig:sdss_range}
\end{figure}

The observability of this signal depends on the orientation of the wake with respect to the observer, i.e. how much of the wake the observer is looking through. By rotating our simulated filaments about an axis parallel to the plane of the wake we can simulate viewing the wake from edge-on (maximally visible) to face-on (minimally visible). After rotating the filaments, the box is re-formed to account for the periodic boundaries. Using the rotated filaments we can determine the viewing angles at which the signal persists (has at least one slice with a $10\%$ density excess over the no wake $1\sigma$ region). The average of the two possible rotations is shown in Figure \ref{fig:slice_angle}, which indicates that for the highest tension string wake the signal in the SCMS+ filaments persists up to an angle of $\sim65$ degrees at $z=10$, $\sim30$ degrees at $z=5$, $\sim30$ degrees at $z=2$, and $\sim15$ degrees $z = 0$. This is in contrast to the visibility of the overdensity of halos in the wake, which is robust to $\sim8$ degrees at higher redshift and $\sim4$ degrees at lower redshift.

\subsection{SDSS Filament Catalog} \label{observed}

We use the publicly available filament catalog from \cite{2016MNRAS.461.3896C} to construct observed slice overdensities to compare to our simulated result (Figure \ref{fig:slice_tension_scms_w5}). The main takeaway of this section is to show that this can be done in a relatively straight-forward way for any catalog of filaments.

The catalog from \cite{2016MNRAS.461.3896C} identified filaments in a combined DR7/DR12 SDSS dataset using SCMS. The filament catalog is a set of points reported in right ascension (RA), declination, and redshift in $\Delta z = 0.005$ bins from $z=0.005$ to $z=0.7$. The smoothing length used in the SCMS method for each redshift bin is also available. For filaments in each redshift bin, we convert the coordinates to Cartesian \cite{astropy:2013, astropy:2018}, then separate the filaments using the method outlined in Section \ref{scms}, using the reported smoothing lengths as a guide to the separation length. We then count the filaments in $10$ degree slices of RA for each redshift bin.

Figure \ref{fig:sdss_range} shows the slice overdensities, grouped into $\Delta z = 0.1$ bins (brown), as well as the overdensities for the entire set of filaments (black). The average of the whole filament set we treat as the no-wake case since cosmic string wakes will only affect a fraction of this volume. Also, shown are the $1\sigma$ and $2\sigma$ deviations from the full set of filaments. We compare the whole filament set to each individual slice in $\Delta z = 0.1$ to determine whether an overdensity similar to what we see in Figure \ref{fig:slice_tension_scms_w5} appears in the observed filament population. Figure \ref{fig:sdss_range} shows that almost all slices fall within the $1\sigma$ range around the total filament density using all SDSS galaxies. This suggests that a cosmic string wake formed with the largest tension, $G\mu = 5\times10^{-6}$, is unlikely except for a small region of parameter space where the wake is parallel to the line of sight.

\subsection{WFIRST Projection} \label{wfirst}

Future surveys, such as WFIRST \cite{2013arXiv1305.5425S} will provide high redshift ($z > 2$) galaxy populations, which can be used to identify high redshift filament populations. For example, at $z=2$, WFIRST is expected to observe approximately $1000$ galaxies per square degree per $\Delta z$ over its $\sim 2000$ square degree field of view \cite{2015arXiv150303757S}, leading to $\sim2\times10^5$ galaxies (with $\Delta z = 0.1$). Our simulations indicate a median of $\sim 100$ halos per filament at that redshift, leading to $\sim 2000$ filaments around $z=2$. From Figure \ref{fig:slice_tension_scms_w5}, it appears that there is a $\sim20\%$ enhancement (i.e. the maximum ratio between a simulation with a wake to the case without a wake) in the number of filaments around the cosmic string wake for the $G\mu = 5\times10^{-6}$ case. Taking into account the angles at which the effect is still visible, and assuming all orientations are equally likely, WFIRST should be able to detect this signal at the level of $0.2/N_{fil}^{-1/2}\times 60/180 \sim 3\sigma$, or $99\%$ confidence.

We can repeat the same approximation at $z=10$. At this high redshift WFIRST, in the optimistic case, projects finding $\sim1000$ galaxies with the High-Latitude Survey \cite{2015arXiv150303757S}. Our simulations indicate at $z=10$ there are $\sim10$ galaxies per filament, so we could identify as many as $\sim100$ filaments at this redshift. For a tension of $G\mu = 5\times 10^{-6}$, our simulations show an enhancement of a factor of $\sim3$, with a signal that persists up to an observation angle of $\sim65$ degrees. This would be detectable by WFIRST at $z=10$ at $\sim 22 \sigma$. By contrast, a tension of $G\mu = 10^{-6}$ shows an enhancement of $\sim 0.1$  at $\sim25$ degrees. It would thus enhance the filament overdensity by only $0.3 \sigma$ and not be detectable.

Figure \ref{fig:slice_tension_scms_w5} shows that the signal at $G\mu = 5\times 10^{-6}$ increases at higher redshift, as the signal becomes more pronounced with respect to the filament overdensity induced by normal structure formation. However, for lower tensions with $G\mu \leq  10^{-6}$ the increase in filament overdensity from the cosmic string wake is always less than the intrinsic variance that structure formation imparts in the spatial filament overdensity. Lower tensions thus produce a sharply reduced signal and so a floor in the cosmic string tension detectable with filament finding methods. This floor is at $G\mu \approx 10^{-6}$, approximately where the relative magnitude of the cosmic string kick at string passage equals the mean particle velocity from structure formation (see Section \ref{velocity}).

\section{Conclusion} \label{conclusion}

In this work we have run a suite of simulations to demonstrate the effect of a cosmic string wake on structure formation, in particular the distribution of cosmic filaments. We have shown that while simple measures, like the number of filaments, do not distinguish the presence of a cosmic string wake in simulations, the spatial distribution of these filaments can. These potential signals are not competitive with constraints derived from the CMB, however they provide an independent constraint relevant at low redshift. They may thus be a viable channel for detection in future experiments or for constraining models that have a stronger signal at relatively low redshift ($z \sim5$). The most promising signals presented here are comparisons between the filament number density far from and near the cosmic string wake, especially across the plane perpendicular to the wake. Examples of this type of comparison are shown in Figures \ref{fig:od_tension_scms}, \ref{fig:slice_tension_scms_w5}. We found that the overdensity signal persists for angles between the wake and the observer of up to $\sim30$ degrees from $z=5$ down to $z=2$.

We have shown that the spatial distribution of filaments can be used to detect the imprint of a cosmic string wake at tensions of $G\mu \gtrsim 5\times 10^{-6}$ and redshifts $z \gtrsim 2$. Lower tensions of $G\mu \lesssim 10^{-6}$ produce a smaller signal and are not detectable at any redshift from galaxy filaments. We found that filaments constructed from current galaxy catalogs (SDSS) show no sign of the overdensities we see in simulations in which a large string tension wake has been included. However, improved sensitivity is expected with new data from galaxy survey experiments in the coming years, such as WFIRST \cite{2013arXiv1305.5425S} or EUCLID \cite{2018LRR....21....2A}. These surveys will easily detect filaments at $z \sim 2$, and present opportunities to probe cosmic strings with tensions between $G\mu=10^{-6}$ and $G\mu=5\times10^{-6}$. If filament finders can be successfully used at higher redshifts, up to $z \sim 10$, very high significance detections may be achieved.

\acknowledgments \label{acknowledgments}
We thank Yen-Chi Chen and Shirley Ho for useful discussions. YC is supported in part by the US Department of Energy grant DE-SC0008541, and thanks the Kavli Institute for Theoretical Physics (supported by the National Science Foundation under Grant No. NSF PHY-1748958) for support and hospitality while the work was being completed. SB was supported by NSF grant AST-1817256. This material is based upon work supported by the National Science Foundation Graduate Research Fellowship under Grant No. DGE-1326120.

\appendix

\section{Box Size Scaling} \label{boxsize}

\begin{figure}
\centering
\includegraphics[width=0.45\textwidth]{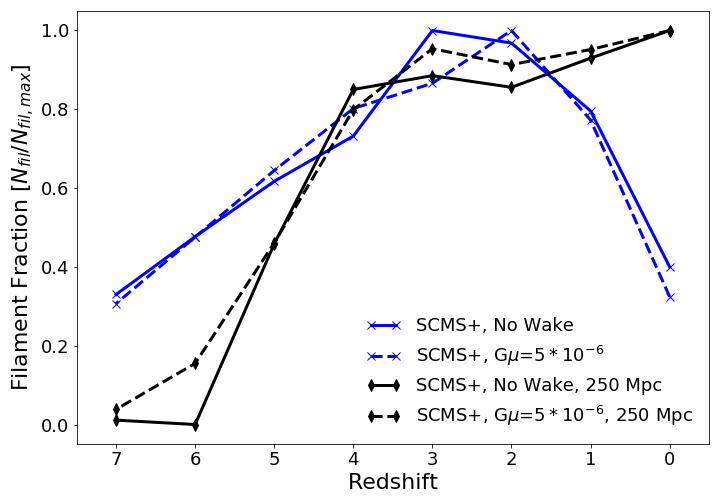}
\caption{\emph{Box size \& resolution check (SCMS+).} Fraction of maximum filaments versus redshift for simulations with different box sizes ($64$ \& $250$ Mpc), but the same structure seed and number of DM particles.}
\label{fig:nfil_boxsize_scms}
\end{figure}

In this appendix we check that our results are not greatly affected by the box size or resolution. We compare one of the main simulations ($64$ Mpc) to a larger box size ($250$ Mpc), lower resolution simulation. Specifically, we look at the SCMS+ filaments from these two simulations, in the case where we have no wake and case where we have the largest tension wake used ($G\mu = 5\times 10^{-6}$). We look only at a single run of the simulation (as opposed to other figures, which are the combination of five runs with different structure seeds). The single simulation means greater variance in the signal, however it is clear from Figure \ref{fig:nfil_boxsize_scms} that while the increased box size has increased the number of filaments, it has not changed the lack of a signal between the no-wake and large-wake simulations (this is true for the DisPerSE filaments as well).\\

\bibliography{refs}

\end{document}